# Electrically tunable valley dynamics in twisted WSe₂/WSe₂ bilayers

Giovanni Scuri[1]*, Trond I. Andersen[1]*, You Zhou[1,2]*, Dominik S. Wild[1], Jiho Sung[1,2], Ryan J. Gelly[1], Damien Bérubé[3], Hoseok Heo[1,2], Linbo Shao[4], Andrew Y. Joe[1], Andrés M. Mier Valdivia[4], Takashi Taniguchi[5], Kenji Watanabe[5], Marko Lončar[4], Philip Kim[1,4], Mikhail D. Lukin[1]† & Hongkun Park[1,2]†

[1]Department of Physics and [2]Department of Chemistry and Chemical Biology, Harvard University, Cambridge, MA 02138, USA

[3]Department of Physics, California Institute of Technology, Pasadena, CA 91125, USA

[4]John A. Paulson School of Engineering and Applied Sciences, Harvard University, Cambridge, MA 02138, USA

[5]National Institute for Materials Science, 1-1 Namiki, Tsukuba 305-0044, Japan

*These authors contributed equally to this work.

†To whom correspondence should be addressed: hongkun_park@harvard.edu, lukin@physics.harvard.edu

**The twist degree of freedom provides a powerful new tool for engineering the electrical and optical properties of van der Waals heterostructures. Here, we show that the twist angle can be used to control the spin-valley properties of transition metal dichalcogenide bilayers by changing the momentum alignment of the valleys in the two layers. Specifically, we observe that the interlayer excitons in twisted WSe₂/WSe₂ bilayers exhibit a high (>60%) degree of circular polarization (DOCP) and long valley lifetimes (>40 ns) at zero electric and magnetic fields. The valley lifetime can be tuned by more than three orders of magnitude via electrostatic doping, enabling switching of the DOCP from ~80% in the *n*-doped regime to <5% in the *p*-doped regime. These results open up new avenues for tunable chiral light-matter interactions, enabling novel device schemes that exploit the valley degree of freedom.**

Valleys represent the crystal momentum states where bands have an extremum [1-6]. Charge carriers or excitons in different valleys can exhibit markedly distinct properties [1], including different spin, optical selection rules, and Berry curvature, leading to a wealth of new physical phenomena such as the valley Hall [7,8] and Nernst effects [9,10]. Moreover, the valley degree of freedom can enable new ways of encoding and processing information beyond traditional schemes based purely on charge [11-13]. The realization of such applications relies on efficiently initializing a large valley polarization and achieving long valley and exciton lifetimes.

Transition metal dichalcogenides (TMDs) are a promising platform for valleytronics as they host tightly bound excitons with coupled spin-valley properties that can be optically addressed via circularly polarized light [2,14-16]. A major obstacle to harnessing these properties in monolayer TMDs is the short lifetimes of intralayer excitons (0.1-1 ps) resulting from the sizeable electron-hole wavefunction overlap [17,18], as well as rapid valley mixing caused by exchange interactions [19-23]. Interlayer excitons in bilayer TMDs [13,24,25], consisting of an electron and a hole residing in two distinct TMD layers, present a promising route for overcoming these limitations. Because of the reduced wavefunction overlap, interlayer excitons can exhibit enhanced lifetimes that are 3 to 4 orders of magnitude longer than their intralayer counterparts [24,25].

Unfortunately, interlayer excitons in naturally occurring TMD bilayers exhibit rapid valley mixing because valleys of opposite chirality are degenerate in energy and momentum (Fig. 1(a), top) [1,26,27]. Moreover, natural bilayers are inversion symmetric (the two layers are rotated 180º relative to each other), thus as a whole they do not exhibit net valley polarization [26,28] unless the symmetry is broken, *e.g.*, with an electric field [29]. Consequently, most valleytronic studies involving interlayer excitons have focused on heterobilayers made of two different materials, such as $WSe_2/MoS_2$ [30] and $MoSe_2/WSe_2$ [12,13,31,32].

In this Letter, we show that the introduction of a twist angle between the layers can provide a new avenue for engineering the spin-valley properties of TMD bilayers, including homobilayers. Previously, the twist angle has been used to modify the resonance energy of excitons [33,34] and alter their properties due to moiré-induced spatial confinement and hybridization of bands [35-40]. Here, we tune the twist angle between the layers to control the momentum alignment of their respective valleys (Fig. 1(a), bottom), permitting long-lived interlayer exciton states with slow valley depolarization even at zero electric and magnetic fields. Importantly, we also demonstrate that the exciton and valley dynamics of twisted bilayers can be tuned via electrostatic doping.

To experimentally demonstrate twist-based spin-valley engineering, we fabricate optically addressable field-effect transistors that incorporate twisted $WSe_2/WSe_2$ bilayers (t-$WSe_2/WSe_2$) encapsulated in hexagonal boron nitride (hBN, Fig. 1(b), Supplementary Section I [41]). The devices feature top and bottom graphene gates for independent control of doping and vertical electric field. Using the tear-and-stack technique [42,43], we fabricate multiple such devices from high-quality exfoliated flakes, with target twist angles ranging from 0º to 17º. For comparison, we also make a device using a 2H-stacked natural bilayer (labeled as 180º twist angle).

Figure 1(c) shows polarization-resolved photoluminescence (PL) spectra from t-$WSe_2/WSe_2$ at different twist angles (see Supplementary Fig. S1 for additional twist angles [41]). These spectra are obtained with both of the graphene gates grounded, so that the TMD layers are intrinsic and under zero vertical electric field. At all twist angles, including the natural 2H bilayer, the spectra show two sets of peaks: the higher energy peaks near 1.7 eV ($X_0$) are assigned to momentum direct (K-K) intralayer transitions [26,28,44], and the lower energy peaks between 1.5 and 1.6 eV ($X_I$) are attributed to interlayer transitions [16,44-46]. This assignment is based on our out-of-plane electric field dependence measurements (Fig. 2(a), Supplementary Section II and Supplementary Fig. S2 [41]), which show a zero (non-zero) linear Stark shift for intralayer (interlayer) excitons, consistent with previous studies [25,44]. Despite their weaker binding energies [25], the interlayer excitons have lower energies than the K-K intralayer excitons, indicating that they do not originate from the momentum direct K-K transition. Instead, the interlayer excitons arise from lower-energy transitions that are momentum indirect, as previously predicted for multilayer TMDs [16].

As the twist angle is increased from 0° to 17°, the interlayer exciton peaks blueshift by almost 80 meV, consistent with reduced interlayer coupling [33,34,47-49] (Fig. 1(c)). In all twisted structures, we observe four or five interlayer exciton peaks separated by 15-17 meV. We note that similar multi-peak structures in twisted MoSe₂/WSe₂ heterostructures have been attributed to the confinement of interlayer excitons in moiré supercells [36]. In t-WSe₂/WSe₂ studied here, the multiple peaks are observed even in natural bilayers (as in Ref. [44]) and their spacing is independent of twist angle, suggesting a different origin. Since their energy separation is similar to the optical phonon energy in WSe₂ [44,50], one possibility is that the peaks are phonon replicas [44]. Further experimental and theoretical studies are necessary to confirm this hypothesis, however.

The spin-valley properties of the interlayer excitons change drastically with the introduction of a twist angle between the two layers, as evidenced by the contrast between co- and cross-polarized emission signals ($I_{co}$ and $I_{cross}$, respectively) in PL measurements. Upon illumination with circularly polarized light, the natural bilayer emits almost equal $I_{co}$ and $I_{cross}$, whereas t-WSe$_2$/WSe$_2$ emits much stronger co-polarized light (Fig. 1(c)). Defining the degree of circular polarization (DOCP) as $\frac{I_{co}-I_{cross}}{I_{co}+I_{cross}}$, we find that while the neutral interlayer exciton DOCP remains close to zero in natural bilayers, it reaches values as high as 60% in twisted bilayers (Fig. 1(d), see also Supplementary Fig. S3 and Supplementary Section III [41]). Such high DOCP suggests much slower valley depolarization dynamics in t-WSe$_2$/WSe$_2$.

Figures 2(b) and (c) show the doping-dependence of PL and DOCP of the 17°-twisted bilayer device (see Supplementary Fig. S4 for similar behavior at a twist angle of 2° [41]), obtained by applying equal voltages to the top and bottom graphene gates to eliminate vertical electric field. While the interlayer exciton DOCP exceeds 80% in the *n*-doped regime, it can be switched to almost zero (<5%) in the *p*-doped regime (Fig. 2(c)). The DOCP stays relatively constant within each of the doping regimes and switches abruptly at their boundaries (Fig. 2(c)). As in previous studies, the intralayer excitons do not exhibit this strong asymmetry [26].

Time-resolved PL measurements of the 17° twisted bilayer sample provide further insight into the exciton and valley dynamics (Figs. 3(a)-(c), corresponding DOCP in Fig. 3(d)). In the intrinsic regime, we fit the data with a biexponential decay [51] convoluted with the instrument response (dotted lines in Figs. 3(a) and (d)), and extract fast and slow time scales of $\tau_1$=0.1 ns and $\tau_2$=0.9 (1.0) ns for co- (cross-) polarized emission. Long interlayer exciton lifetimes are also observed for twist angles of 0° and 5°, which exhibit $\tau_2 = 2.0$ ns and $\tau_2 = 3.3$ ns, respectively (Supplementary Fig. S5 [41]). Similar to previous studies [23,52], we observe a 15 ps delay between the co- and cross-polarized emission, which causes a dip in the DOCP until both polarization branches reach the slow decay regime (Fig. 3(d)). Fitting the DOCP in this regime with a linear coupled model (see Supplementary Section IV [41]), we extract a very long valley lifetime of 44 ns (dashed line in inset of Fig. 3(d), lines corresponding to 5 ns and 10 ns shown for comparison), comparable to the best reported values in TMD heterobilayers [31].

In the doped cases, we focus on shorter time scales, because the exciton population and DOCP decay very rapidly in the $n$- and $p$-doped regimes, respectively. Fitting the data with the linear coupled model [41,52] (dashed lines in Figs. 3(b)-(d)), we find that the initial exciton decay rate is faster in the $n$-doped regime ($\tau_1 = 40$ ps) than in the intrinsic and $p$-doped regimes ($\tau_1 = 0.1$ ns). Conversely, the valley lifetime is longer in the $n$-doped regime ($\tau_v = 2.2$ ns) than in the $p$-doped regime ($\tau_v = 30$ ps, similar to instrument response time).

The observed polarization properties and their doping dependence can be understood from the band structure of t-WSe$_2$/WSe$_2$ bilayers. In natural (2H) bilayer WSe$_2$, recent angle-resolved photoemission spectroscopy measurements [53,54] and density functional theory calculations [44,55] suggest that the conduction band minimum (CBM) and valence band maximum (VBM) are located at the Q and K points, respectively. Our results suggest that this is also the case in t-WSe$_2$/WSe$_2$. In particular, the electric field dependent PL measurements display an interlayer exciton Stark shift that corresponds to an electron-hole separation of $d = 0.37$ nm (Fig. 2(a)). This value is smaller than the interlayer separation of $d_0 = 0.6$ nm [25] but larger than $d_0/2$ as expected for a K (hole) to Q (electron) transition, where the hole is localized in a single layer and the electron is partially delocalized between the layers (see Supplementary Section II for further discussion [41]).

A key difference in the band structures of natural and twisted WSe$_2$ bilayers is that the Q and K points are spin-degenerate in the former, but not in the latter [56]. Therefore, electrons and holes in natural bilayer WSe$_2$ do not need to acquire any energy or momentum to change spin, enabling rapid depolarization (Fig. 4(a)). In contrast, valley depolarization of neutral interlayer excitons in twisted structures requires both carriers to scatter with phonons, acquire energy and flip their spins [23,51] (Fig. 4(b)). This is a much slower process, and the dynamics are instead likely governed by electron-hole exchange interactions [12,51,57]. However, this process is also slow, because the interlayer excitons are indirect both in real and momentum space. In our experiment, we show that the valley depolarization rate is much slower than the exciton decay rate (Fig. 3), leading to the large observed DOCP (Fig. 1(d)).

The observed hierarchy of valley lifetimes in the three doping regimes, *i.e.* $\tau_v$(intrinsic) > $\tau_v$($n$-doped) > $\tau_v$($p$-doped), is also well described by our band-structure considerations. In the $p$- and $n$-doped regimes, the depolarization dynamics of charged interlayer excitons are dominated by intervalley scattering [12,51,57], and in contrast to the intrinsic regime, only one of the carriers needs to scatter, due to the additional resident charge in the opposite valley. Positively charged excitons, which already have resident holes in both the K and K' valleys, only require scattering of the electron to depolarize (Fig. 4(c)). Conversely, only the hole needs to scatter in negatively charged excitons (Fig. 4(d)). These considerations indicate that the valley lifetime should be shorter in the doped regimes than in the intrinsic regime, as observed here. Our results in Figure 3(d) suggest that the depolarization mechanism for holes is slower than for electrons in our twisted structures, consistent with the large spin-orbit coupling at the K point VBM and the resultant strong spin-valley locking for holes [26].

The strong asymmetry in degree of circular polarization shown in Figure 2 is a result of the interplay between the depolarization and exciton decay dynamics. In the hole doped regime, the depolarization timescale is shorter than $\tau_1$ (30 ps and 0.1 ns, respectively), resulting in a low DOCP (<5%). In the electron doped regime, on the other hand, $\tau_v$ is much longer than $\tau_1$ (2.2 ns and 40 ps, respectively) so the observed DOCP is very large (>80%).

In conclusion, we have shown that the twist angle provides a new route for engineering the chiral optical properties of interlayer excitons in $WSe_2/WSe_2$ bilayers. Specifically, by changing the momentum separation of the valleys in the two layers, we achieved large valley polarization for long-lived interlayer excitons. Such valley polarization, which can be initialized and read out optically, is a direct consequence of the valley lifetime greatly exceeding the exciton lifetime. Moreover, in agreement with band structure considerations, the valley dynamics are highly tunable by electrostatic doping, enabling switching of DOCP from more than 80% in the *n*-doped regime to <5% in the *p*-doped regime. Our results exemplify the power of twist-based engineering in van der Waals heterostructures and demonstrate that twisted TMD bilayers are a new promising platform for tunable chiral photonics and electrically switchable spin-valley based devices. The coupling to functional structures, such as plasmonic metasurfaces [58-60], can further enhance chiral light-matter interactions and enable routing of electrically switchable chiral photons, thus opening up avenues for applications in quantum information storage and processing.

We acknowledge support from the DoD Vannevar Bush Faculty Fellowship (N00014-16-1-2825 for H.P., N00014-18-1-2877 for P.K.), NSF (PHY-1506284 for H.P. and M.D.L.), NSF CUA (PHY-1125846 for H.P. and M.D.L.), AFOSR MURI (FA9550-17-1-0002), AFOSR DURIP (FA9550-09-1-0042), ARL (W911NF1520067 for H.P. and M.D.L.), the Gordon and Betty Moore Foundation (GBMF4543 for P.K.), ONR MURI (N00014-15-1-2761 for P.K.), and Samsung Electronics (for P.K. and H.P.). All fabrication was performed at the Center for Nanoscale Systems (CNS), a member of the National Nanotechnology Coordinated Infrastructure Network (NNCI), which is supported by the National Science Foundation under NSF award 1541959. K.W. and T.T. acknowledge support from the Elemental Strategy Initiative conducted by the MEXT, Japan and the CREST (JPMJCR15F3), JST.

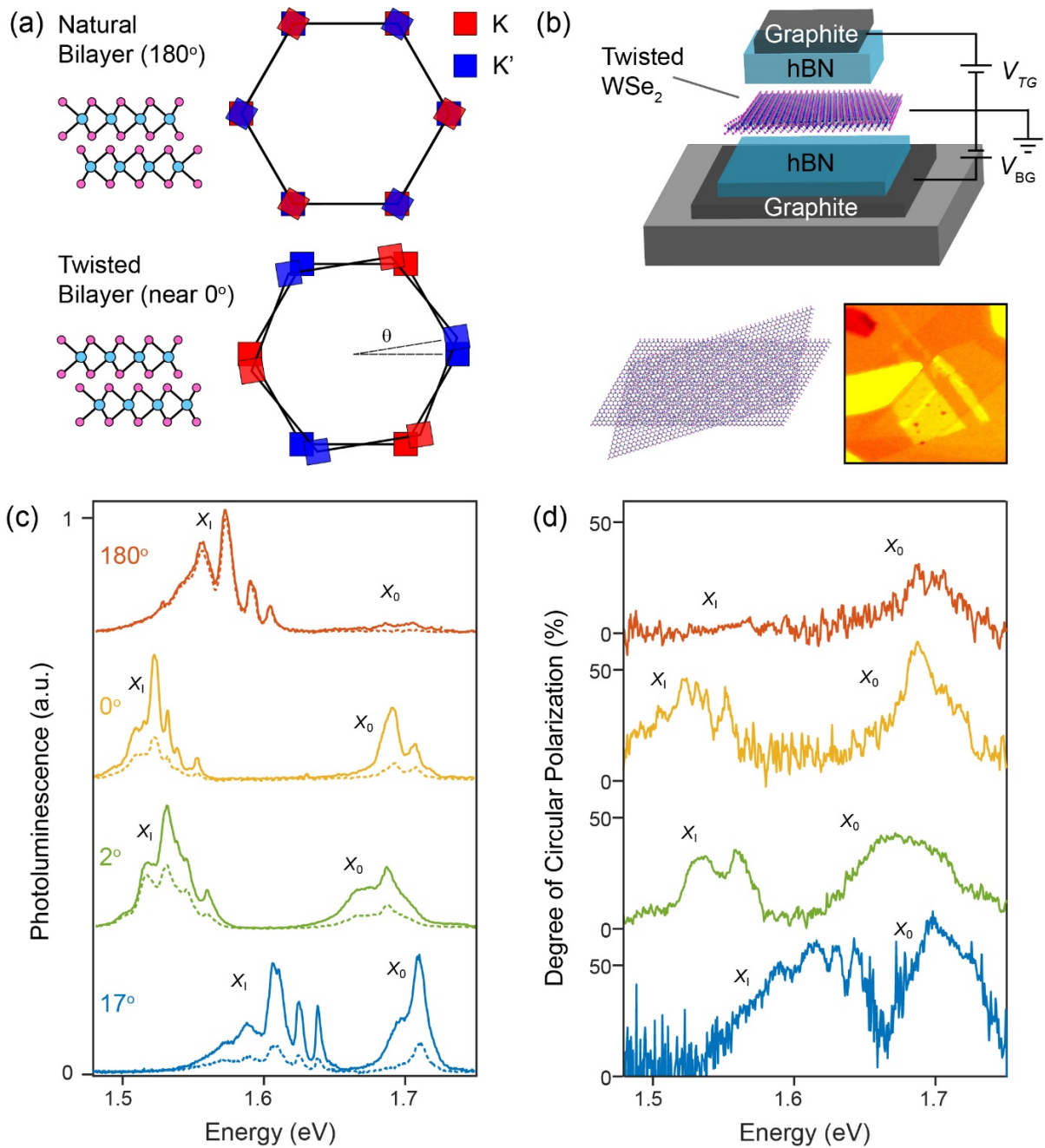

**Fig. 1: Band engineering through twisting:** (a) Side view and Brillouin zone alignment of natural (2H-stacked, top) and twisted homobilayer (bottom). The K and K' points are aligned in natural bilayers, but not in twisted structures. (b) Device schematic (top), illustration of a moiré pattern in a twisted WSe$_2$/WSe$_2$ bilayer (bottom left) and optical image of a device (bottom

right). (c) Polarization-resolved PL spectra from bilayers with varying twist angle. All spectra are collected in the intrinsic regime. Solid (dashed) lines represent co- (cross-) polarized emission. $X_0$ and $X_1$ indicate intra- and interlayer excitons, respectively. (d) Degree of circular polarization calculated from PL spectra in (c). While the natural bilayer (red) exhibits almost zero interlayer DOCP, the twisted devices show DOCP as high as 60%.

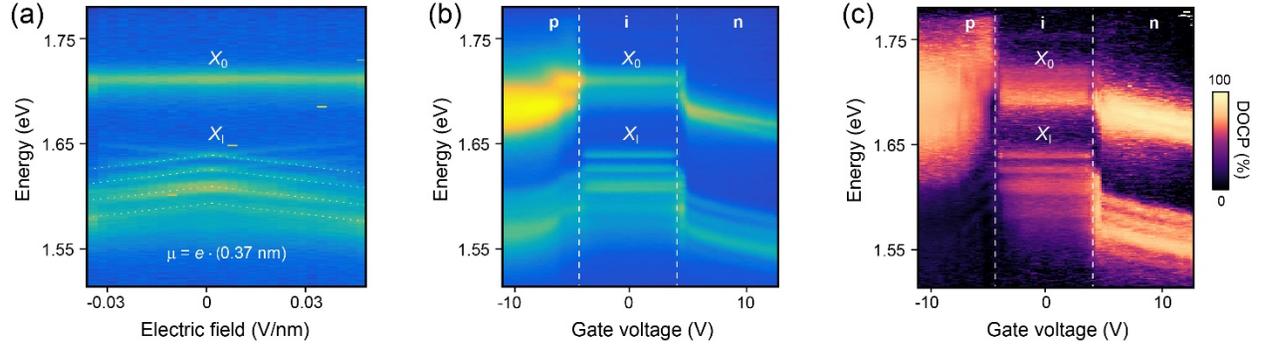

**Fig. 2: Gate tunability of DOCP.** (a) Out-of-plane electric field dependence of intra- and interlayer exciton PL from the 17°-twisted bilayer. While the intralayer excitons ($X_0$) do not shift with electric field, the interlayer excitons ($X_I$) exhibit a Stark shift equivalent to an electron-hole separation of 0.37 nm. Opposite voltages are applied to the two gates (the hBN flakes have the same thickness) to apply an electric field while keeping the TMD intrinsic. (b) Doping dependence of PL emission for intra- and interlayer excitons. The same voltages are applied to the two gates to dope the sample without applying an electric field. (c) Doping dependence of DOCP, showing switching from large (>80%) values in the *n*-doped regime, to almost zero (<5%) in the *p*-doped regime.

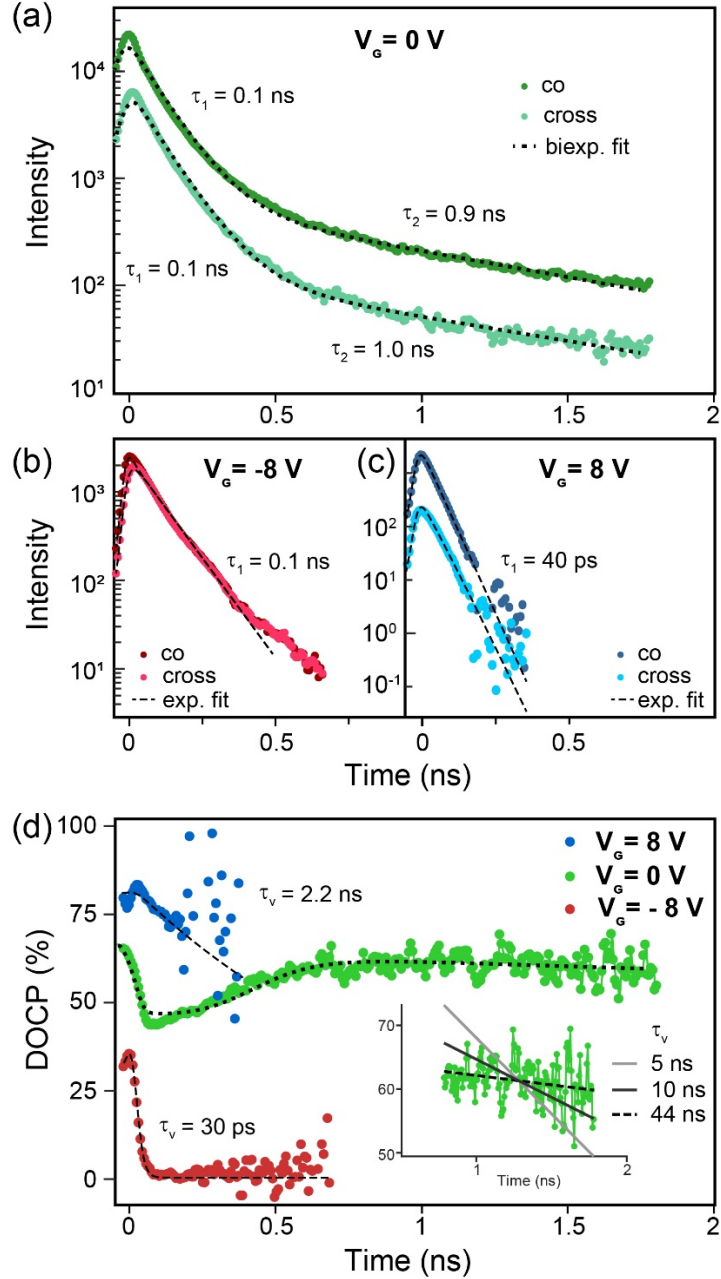

**Fig. 3: Exciton and valley dynamics in twisted WSe₂.** (a) Time-resolved measurements of co- and cross-polarized photoluminescence (dark and light green markers, respectively) from the 17° twisted bilayer in the intrinsic regime. Black dotted lines are bi-exponential fits convolved with the instrument response (Supplementary Section IV [41]). The extracted slow and fast time scales are $\tau_1$=0.1 ns and $\tau_2$=0.9 (1.0) ns for co- (cross-) polarized emission. (b) and (c) Same as in (a), but at $V_G$=-8 V and $V_G$=8 V. Black dashed lines are fits based on the coupled linear model (Supplementary Section IV [41]). The extracted exciton and valley lifetimes are $\tau_1$=0.1 ns and $\tau_v$=30 ps at $V_G$=-8 V, and $\tau_1$=40 ps and $\tau_v$=2.2 ns at $V_G$=8 V. (d) Time-resolved DOCP corresponding to measurements in (a)-(c). Black dotted and dashed curves are calculated from fits in (a)-(c). Inset: Zoomed-in version of long tail in intrinsic regime, fitted with the linear coupled model (dashed line). The extracted valley lifetime is $\tau_v$=44 ns. Lines corresponding to $\tau_v$=5 ns (light grey) and $\tau_v$=10 ns (dark grey) are shown for comparison.

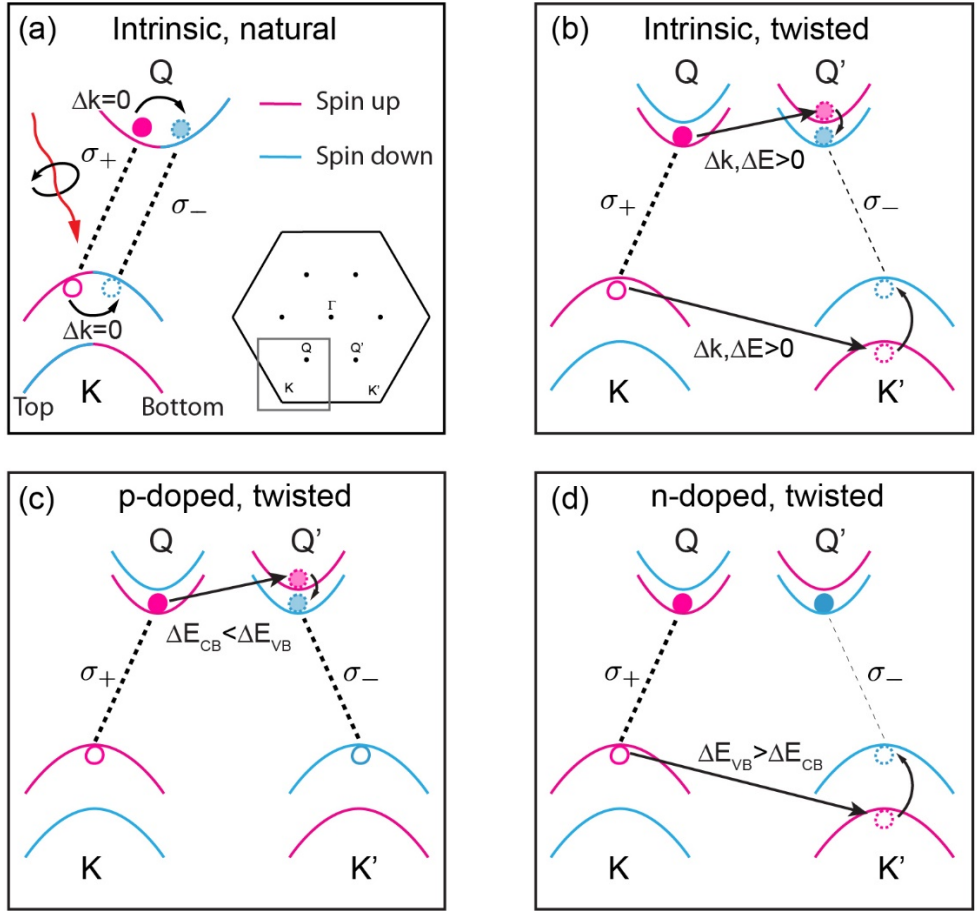

**Fig. 4: Depolarization mechanisms in natural and twisted WSe₂/WSe₂ bilayers.** (a) Depolarization in natural bilayer WSe₂ occurs simply through two spin-flip transitions (curved arrows), causing low DOCP. Magenta (cyan) bands indicate spin up (down) for electrons. The thickness of the dashed lines qualitatively represents the relative dominance of the $\sigma_+$ and $\sigma_-$ recombination processes. Inset: Brillouin zone showing location of the degenerate K and Q points of a natural bilayer as a whole. (b) Depolarization in twisted bilayer WSe₂ requires both a phonon-induced momentum kick (straight arrows) and a spin-flip transition (curved arrows) for both carriers, resulting in high DOCP and long $\tau_v$. (c)-(d) Since positively (negatively) charged excitons already have holes (electrons) at both the K and K' (Q and Q') points, only the electron (hole) must flip spin and shift in *k*-space.

# Supplemental Material for: Electrically tunable valley dynamics in twisted WSe₂/WSe₂ bilayers

Giovanni Scuri*, Trond I. Andersen*, You Zhou*, Dominik S. Wild, Jiho Sung, Ryan J. Gelly, Damien Bérubé, Hoseok Heo, Linbo Shao, Andrew Y. Joe, Andrés M. Mier Valdivia, Takashi Taniguchi, Kenji Watanabe, Marko Lončar, Philip Kim, Mikhail D. Lukin† & Hongkun Park†

*These authors contributed equally to this work.

†To whom correspondence should be addressed: hongkun_park@harvard.edu, lukin@physics.harvard.edu

# I. MATERIALS AND METHODS

Flakes of hBN, graphene and $WSe_2$ were mechanically exfoliated from bulk crystals onto Si wafers (with 285 nm $SiO_2$). Monolayer and bilayer $WSe_2$ were identified using optical microscopy and later confirmed in photoluminescence measurements. The thicknesses of hBN flakes were determined with atomic force microscopy. The heterostructures were then assembled with the dry-transfer method, using the tear-and-stack technique [42,43] to form twisted bilayer $WSe_2$. Next, electrical contacts to the $WSe_2$ and graphite gates were defined with electron-beam lithography and deposited through thermal evaporation (10 nm Cr+90 nm Au). Optical measurements were conducted in a 4 K cryostat (Montana Instruments), using a self-built confocal setup with an 0.75 NA objective. The DOCP was measured with two polarizers in the excitation and collection paths, and a quarter wave plate directly above the objective. To eliminate any polarization-dependent instrument response, all DOCP measurements included excitation with (and collection of) both $\sigma_+$ and $\sigma_-$ light. PL measurements were carried out using a 660 nm diode laser. A sub-picosecond pulsed laser (Coherent) and a streak camera (Hamamatsu) were used in the time-resolved PL measurements.

# II. EXCITON CHARACTERIZATION BASED ON DC STARK SHIFT

Upon applying a vertical electric field, excitons experience a Stark shift given by $\Delta E = E \cdot e \cdot d$, where $E$ is the applied electric field, $d$ is the (vertical) electron-hole separation, and $e$ is the elementary charge. Since interlayer excitons have an out-of-plane dipole moment ($d$), unlike intralayer excitons, the two species can be distinguished through electric field dependent PL measurements (Fig. 2(a) and Fig. S2). In all of our devices, the higher-energy feature ($\sim 1.7$ eV) exhibits no Stark shift, and is therefore attributed to intralayer excitons. The lower-energy peaks, on the other hand, show a clear shift, and are therefore assigned to interlayer excitons.

The interlayer excitons are expected to be momentum indirect, because their PL energy is lower than that of the direct intralayer exciton, even with their weaker binding energy [25]. This is further supported by the fact that the extracted electron-hole separation, $d = 0.37$ nm (0.36 nm in 2° twist device), is significantly smaller than the full interlayer separation ($d_0 = 0.6$ nm) [25] that would be expected if both carriers were at the K point. It is also not consistent with the Γ-K transition, because the wavefunction at the Γ-point is completely delocalized between the two layers [33,61], causing $d \leq d_0/2$. Instead, the extracted electron-hole separation is consistent with the K to Q transition, where the hole is localized in a single layer and the electron is partially delocalized.

## III. SPATIAL DEPENDENCE OF DOCP IN A SAMPLE CONTAINING BOTH NATURAL AND TWISTED BILAYER REGIONS

To confirm the stark contrast in DOCP between natural and twisted bilayer WSe$_2$, we fabricate an hBN-encapsulated TMD heterostructure that contains both natural (180°) and twisted (~0°) bilayer regions (Fig. S3(a)). The device was made from a single exfoliated WSe$_2$ flake that had both a bilayer and a monolayer region, the latter of which was torn and stacked on top of itself. Integrating only the interlayer exciton emission (photon energies below 1.6 eV), we find that the natural bilayer area exhibits almost no DOCP, while the twisted region has a DOCP close to 50% almost everywhere, except in sporadic defect spots (Fig. S3(b)).

## IV. FITTING OF LIFETIME DATA

To fit the full time-dependence of the photoluminescence in the intrinsic regime, we use a biexponential decay convoluted with the system response, $s(t)$, as measured from the response of our sub-picosecond laser:

$$p(t) = \left(A_1 e^{-t/\tau_1} + A_2 e^{-t/\tau_2}\right) * s(t - t_0).$$

Here, $\tau_1$ and $\tau_2$ are the fast and slow decay timescales, with corresponding amplitudes $A_1$ and $A_2$. This model is used for both the co- and cross-polarized emission components (with separate fit parameters), and $t_0$ accounts for the observed delay between the two. The fits are shown as dotted lines in Fig. 3(a), with corresponding DOCP in Fig. 3(d).

In the $n$- and $p$-doped regimes, we focus on shorter timescales due to their shorter exciton and valley lifetimes, respectively. At these timescales, we only observe a single exponential decay and therefore fit with a linear coupled model, which also allows for extracting the valley lifetime, $\tau_v$:

$$\dot{p}_{\text{co}} = -\frac{p_{\text{co}}}{\tau_1} - \frac{p_{\text{co}} - p_{\text{cross}}}{\tau_v} + (1 - \alpha) \cdot s(t),$$

$$\dot{p}_{\text{cross}} = -\frac{p_{\text{cross}}}{\tau_1} - \frac{p_{\text{cross}} - p_{\text{co}}}{\tau_v} + \alpha \cdot s(t).$$

Here, $p_{\text{co}}$ and $p_{\text{cross}}$ are the co- and cross-polarized exciton populations, and $\alpha$ is the proportion of cross-polarized excitons due to imperfect excitation. In order to account for the system response, we use the measured laser pulse, $s(t)$, as the laser input. The fits and corresponding DOCP are shown with dashed lines in Fig. 3(b)-(d). This model was also used for fitting in the intrinsic regime at longer timescales (inset of Fig. 3(d)).

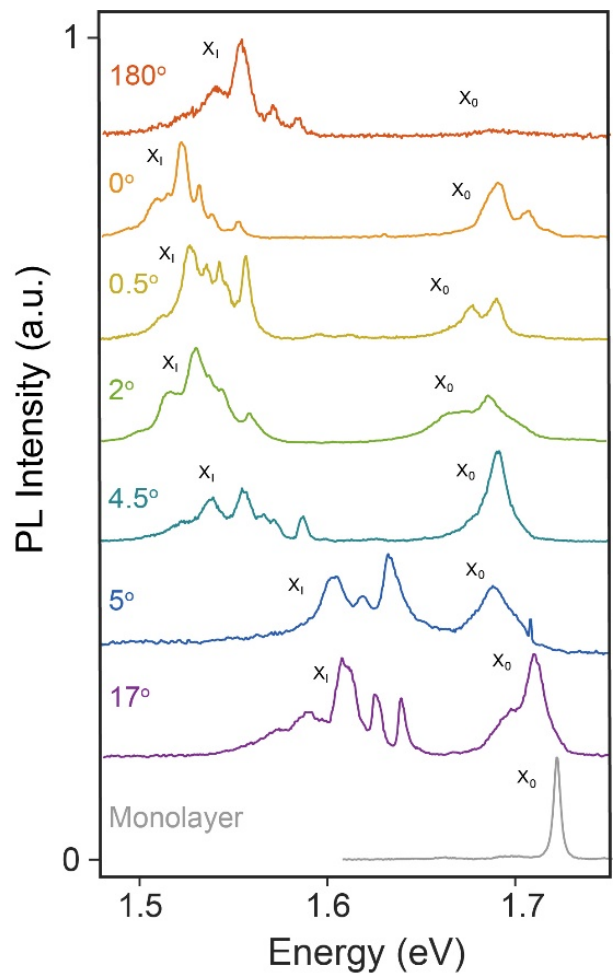

**Fig. S1: Photoluminescence from additional samples.** Colored (grey) curves show PL from bilayer (monolayer) WSe$_2$, including additional twist angles of 0.5º, 4.5º and 5º not displayed in the main text. The devices with 4.5º and 5º twist angles show the same interlayer exciton blue-shift as was presented for the 17º sample in the main text.

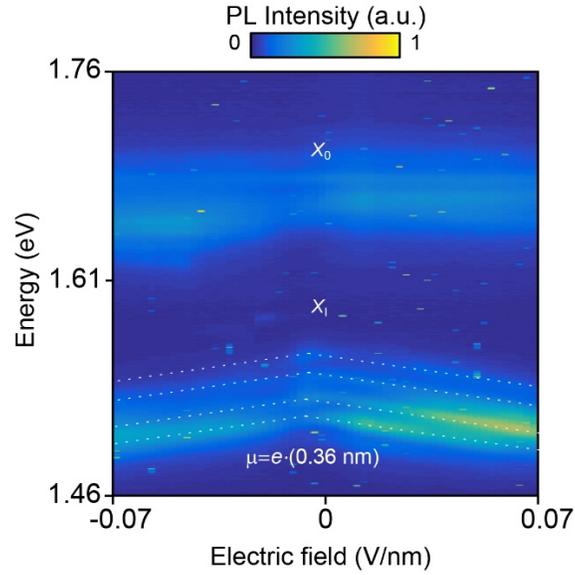

**Fig. S2: Electric field dependence of PL in device with 2º twist angle.** While the intralayer exciton ($\sim 1.7$ eV) does not shift observably with electric field, the interlayer excitons ($\sim 1.5 - 1.6$ eV) exhibit a clear Stark shift (dashed white lines). The corresponding electron-hole separation is 0.36 nm, similar to the value extracted in the 17º device (0.37 nm).

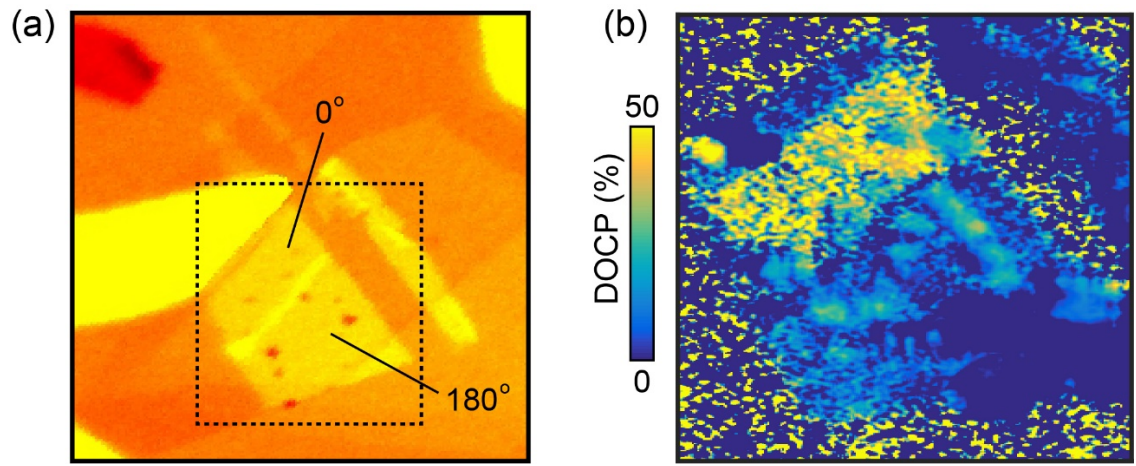

**Fig. S3: Spatial map of DOCP. (a)** Optical image of an hBN-encapsulated heterostructure that contains both natural (180°, lower right) and twisted (~0°, upper left) bilayer regions. (b) Spatial dependence of average interlayer exciton DOCP (photon energy below 1.6 eV). Consistent with PL spectra shown in the main text, the twisted bilayer region exhibits much higher DOCP than the natural bilayer region.

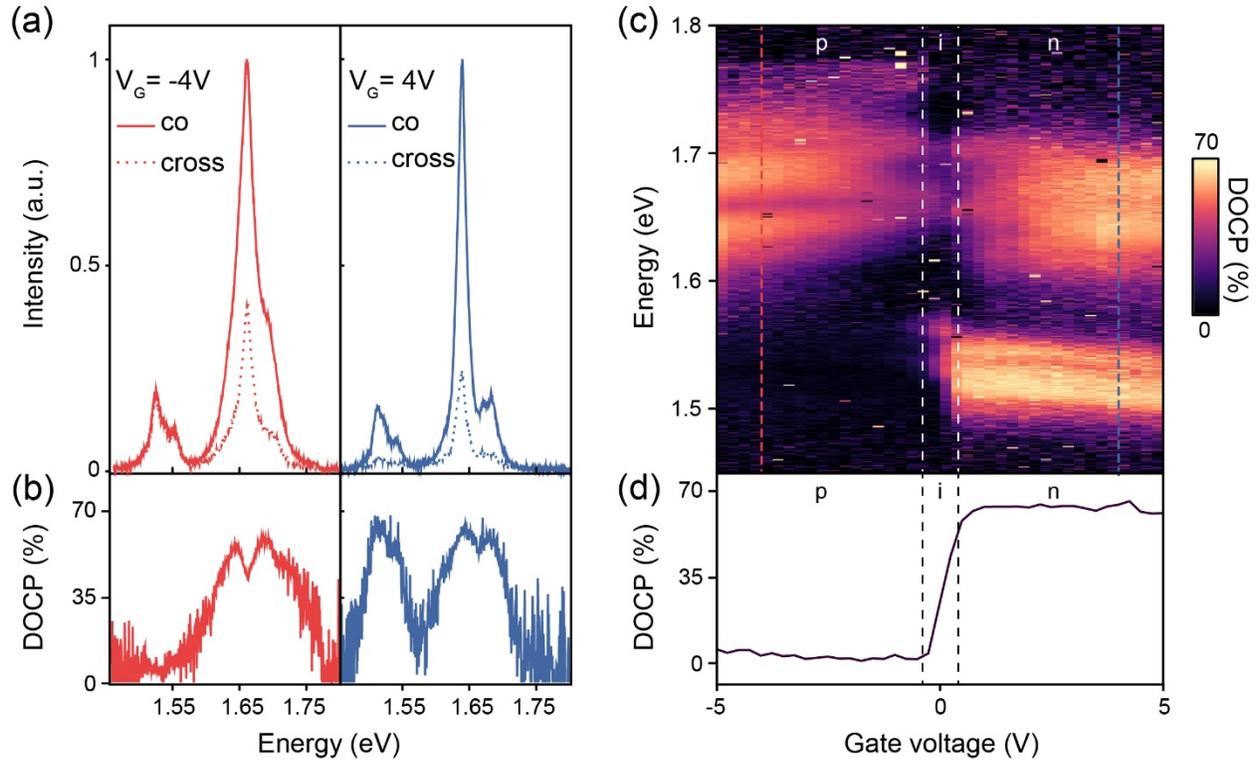

**Fig. S4: Doping dependence of DOCP in device with 2° twist angle.** (a) Polarization-resolved photoluminescence spectra from 2°-twisted bilayer WSe$_2$ at gate voltages $V_G$=-4 V (left) and $V_G$=4 V (right). Solid and dashed curves show co- and cross-polarized emission, respectively. (b) DOCP calculated from PL spectra in (a). As in the 17° device presented in the main text, the interlayer DOCP is much higher in the *n*-doped regime than in the *p*-doped regime. (c) Full gate dependence of DOCP. (d) DOCP of the brightest interlayer exciton peak as a function of gate voltage. Since this device exits the intrinsic regime at relatively low gate voltages, plateaus are only observed in the *p*- and *n*-doped regimes.

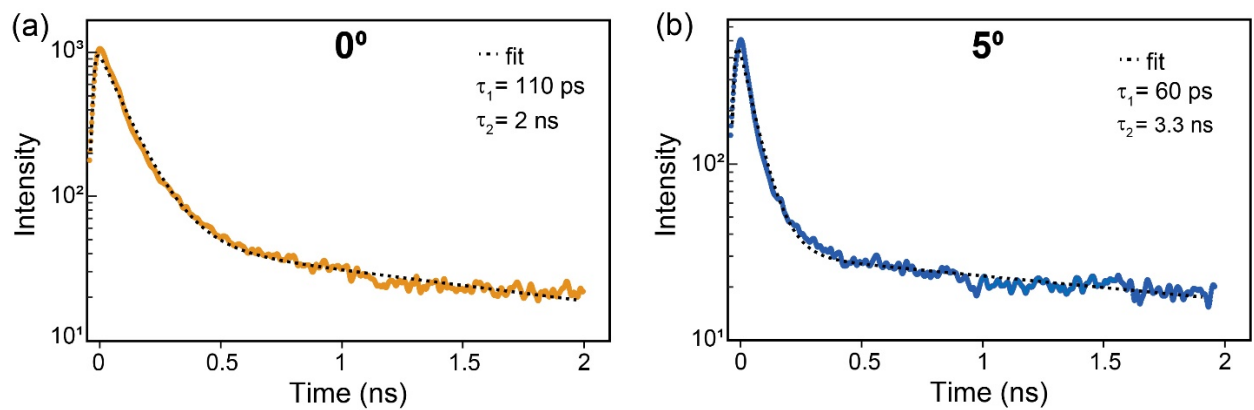

**Fig. S5: Exciton lifetime in additional samples.** (a)-(b) Time-resolved PL measurements from devices with 0º (a) and 5º (b) twist angles (colored dots). Dotted black lines are biexponential fits convoluted with the system response. The extracted fast and short timescales are $\tau_1$=110 (60) ps and $\tau_2$=2.0 (3.3) ns for the device with 0º (5º) twist angle.